# Opportunities for Analog Coding in Emerging Memory Systems

Jesse H. Engel[1,2,5,*], S. Burc Eryilmaz[2], SangBum Kim[3], Matthew BrightSky[3], Chung Lam[3], Hsiang-Lan Lung[4], Bruno A. Olshausen[1], and H.-S. Philip Wong[2]

[1] Redwood Center for Theoretical Neuroscience, UC Berkeley, Berkeley, CA, 94720
[2] Dept. of Electrical Engineering and Stanford SystemX Alliance, Stanford University, Stanford, CA, 94305
[3] IBM Research, T.J. Watson Research Center, Yorktown Heights, NY, 10598
[4] Macronix International Co., Ltd., Emerging Central Lab, 16 Li-Hsin Road, Hsinchu Science Park, Taiwan
[5] Google Brain, 1965 Charleston Rd., Mountain View, CA, 94043
[*] Email: jesse.engel@gmail.com

**Abstract**
------------
The exponential growth in data generation and large-scale data analysis creates an unprecedented need for inexpensive, low-latency, and high-density information storage. This need has motivated significant research into multi-level memory systems that can store multiple bits of information per device. Although both the memory state of these devices and much of the data they store are intrinsically analog-valued, both are quantized for use with digital systems and discrete error correcting codes. Using phase change memory as a prototypical multi-level storage technology, we herein demonstrate that analog-valued devices can achieve higher capacities when paired with analog codes. Further, we find that storing analog signals directly through joint-coding can achieve low distortion with reduced coding complexity. By jointly optimizing for signal statistics, device statistics, and a distortion metric, finite-length analog encodings can perform comparable to digital systems with asymptotically infinite large encodings. These results show that end-to-end analog memory systems have not only the potential to reach higher storage capacities than discrete systems, but also to significantly lower coding complexity, leading to faster and more energy efficient storage.

**Introduction**
-------------------

A paradigm shift in the type and quantity of storable information is currently under way. Internet connected devices are projected to dramatically increase in number to 50 billion, more than 6 per a person, by the year 2020 [1]. Much of the data produced by these devices, such as pixel intensities from cameras, sound recordings from microphones, and time-series from sensors, will come from signals that are intrinsically analog-valued or many-valued (>100 values). These signals are also highly redundant and compressible, with a large degree of the variance explained by a smaller number of factors than the intrinsic dimensionality of the signal.

Concurrently, a second paradigm shift is under way in the media we use to store data. Early storage media such as phonograph records and VCR tapes relied on perturbing an analog-valued state (wax height and magnetic polarization, respectively). Digital computation led to the popularity of binary storage representations that inhibit noise propagation and utilize the concurrently developed theories of binary error correcting codes [2]. However, many emerging memory technologies have shifted back to analog-valued media to create multi-level devices that fill the need for inexpensive and high-density storage. MLC-Flash, Phase Change Memory (PCM), Resistance RAM (RRAM), and Conductive Bridge RAM (CBRAM) are all examples of technologies that have analog state (threshold voltage or resistance) determined by the gate-charge, resistive amorphous capping region, and conducting filament respectively [3, 4].

Despite the prevalence of digital representations in current technology, we examine herein the potential for memory systems that utilize analog representations (analog-valued signals stored in analog-valued devices) to achieve higher capacity, reduce latency, and reduce coding complexity.

To create a fair comparison between digital and analog systems, we need to compare their end-to-end performance on a storage task. The rate-distortion theory introduced by Shannon in 1948, provides a common framework to evaluate both approaches [5]. As diagramed for joint source-channel coding in Fig. 1a, a source signal $S^k$, of dimensionality $k$, is encoded ($F$) into a representation $V^m$, of dimensionality $m$. The encoded signal is then stored in memory devices, equivalent in the communications literature to passing through a noisy channel ($P(R|V)$), resulting in a noisy recalled signal $R^m$. A decoder ($G$) then creates an estimate of the original source ($\hat{S}^k$). For the PCM devices presented here, $V$ and $R$ directly correspond to the write voltages and read resistances of $m$ devices used to store $k$ symbols of data.

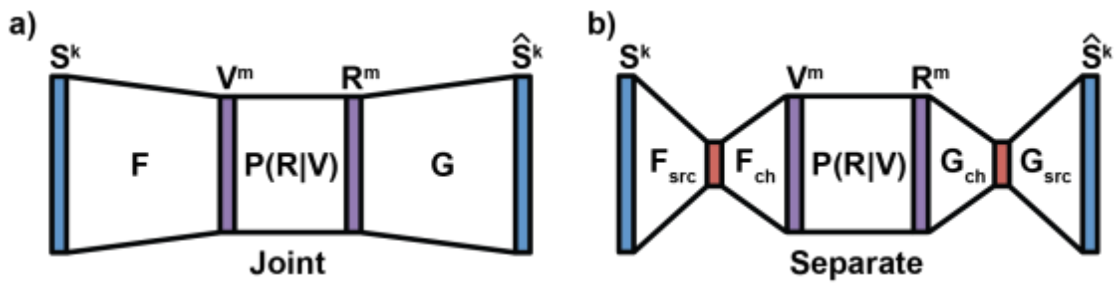

*Figure 1: Illustration of Coding Strategies.* The dimensionality of signal is represented by the size of the bar. *(a) Joint Coding.* A k dimensional source, $S^k$, is encoded with a function F into an m dimensional representation, $V^m$. This is stored into analog devices, and later read as the resistances $R^m$. These resistances are decoded with the function G into a reconstructed source signal ($\hat{S}^k$). *(b) Separate Coding.* Each encoding and decoding step is broken into separate source and channel procedures ($F_{src}$, $F_{ch}$ and $G_{src}$, $G_{ch}$ respectively). In discrete systems, source coding is usually applied in software (such as JPEG for images), and channel coding is applied in an on-chip memory controller. Such schemes can be optimal, but require asymptotically large blocklength to do so (k and m go to infinity, while k/m stays constant).

The distortion (D) is determined by comparing the source and the reconstructed source with a *task-relevant* metric. For example, if the mean squared error (MSE) is a relevant metric, $D = \frac{1}{N}\sum_{i=0}^{N}(S_i - \widehat{S_i})^2$, where N is the total number of examples. For media such as images, perceptual distortion metrics such as mean structural similarity index (MSSIM) may be more appropriate [6]. For a given source, the rate (R, not to be confused with resistance) is given by the compression ratio k/m. For example, in a binary system with rate R, a source signal of k bits can be stored in m bits of memory and reliably recalled with a distortion D.

As shown in Fig. 1b, encoding and decoding can each be divided into separate coding procedures for the source and the channel ($F_{src}$, $F_{ch}$, $G_{src}$, $G_{ch}$). In this separate coding paradigm, source coding aims to remove statistical redundancy from the data. The resulting compressed representation requires less resources to store, but is extremely fragile to noise. Consequently, channel coding introduces redundancy and statistical dependencies through an error correcting code (ECC) (typically with addition of parity bits dependent on data bits through a graphical model [7] ), enabling the original source coded data to be decoded/inferred from the noisy version.

Since small errors in a source coded signal can produce large distortions, sufficient redundancy must be added to ensure that no uncorrected errors remain, leaving source coding as the sole source of distortion. The degree of required redundancy (devices/bit) is determined by inverse of the information capacity (C, bits/device) of the storage devices themselves, which is the maximum rate achievable with an optimal ECC. Thus, the capacity of an analog-valued memory device provides a fair means of comparison between digital and analog implementations. For each symbol produced by a source, a lower bound on R(D) (devices/symbol) for a separate-coded digital system is given by R(D) (bits/symbol) of the source coder divided by the capacity of the memory device (bits/device).

Beyond information capacity, Shannon proved that errorless transmission is only assured for separate coding in the limit that the code word length, also know as the block length, approaches infinity [5]. Modern binary ECCs with rates that approach the Shannon capacity, such as turbo codes [8], often do so at the expense of large block lengths (>1kB). While the degree of redundancy remains constant for large block lengths, the expense of the decoder increases dramatically as it must perform inference on problems of higher and higher dimensionality. Even well designed LDPC decoders with large block lengths can consume significant power (~100 mW – 1 W) and area (~ 10s $mm^2$) [9, 10]. As code lengths increase, so do latencies due to decoding, in one case rising from 20us to 220us for an increase in BCH (Bose-Chaudhuri-Hocquenghem) code length from 16B to 2048B [11].

While increasing the dimensionality (k) of the signal improves coding performance, it also requires reading from larger segments of memory at once and increased complexity of decoder circuitry, resulting in larger access latency and more energy consumption. However, in the finite block length regime, joint coding can outperform separate coding schemes [12]. Rather than trying to make the channel behavior asymptotically deterministic, joint coding attempts to map the statistics of the source through the stochasticity of the channel such as to produce a low average distortion.

Under certain conditions, such mappings can achieve optimal R(D) with finite block length. Trivial, but common, examples include the Gaussian source through a Gaussian channel with MSE distortion and the Bernoulli source through a binary symmetric channel with Hamming distance distortion. In both cases, the optimal solution is achieved by performing no coding at all and sending each symbol individually (block length of one) [13]. These examples demonstrate how matching the statistics of the source, channel, and cost can significantly increase the efficiency of reliable communication.

Here, we show that using analog codes with analog memory devices can improve system performance over digital codes in both the separate coding and joint coding regimes. Using PCM as an example analog-valued memory, we demonstrate that separate coding systems can reach higher capacities when paired with analog ECCs. Further, we find that measuring analog device statistics enables efficient memory design of analog systems with tradeoffs of capacity and energy consumption. In the joint coding regime, we develop optimal mappings for storing analog values directly in the analog resistances of the PCM devices. For symbol-by-symbol storage of a Gaussian source with a MSE distortion metric, we find comparable R(D) performance to digital implementations employing optimal capacity-achieving channel coding. This result is achieved with a block length of 1, indicating the potential for better performance with higher dimensional encodings, utilizing correlations in the source statistics. This work strongly motivates further research into analog coding schemes for emerging memories, including the use of higher dimensional joint coding schemes (such as artificial neural networks) and storing high-dimensional redundant source signals (such as natural images and audio).

**Results**
-----------
To measure the capacity of analog-valued PCM devices, we recast the storage problem as a communication problem. As seen in Fig 2a, the devices are perturbed by voltage pulses of different magnitudes corresponding to the input probability distribution P(V). These pulses modulate the device resistances, resulting in an output resistance distribution P(R). The conditional dependence of the read resistance on the write voltage P(R|V), defines the noisy channel through which information must be communicated. In the limit of an infinite block-length code, the capacity is the number of bits of information per use of the device that can be communicated with error [5]. It is equivalently given by the maximum mutual information between the input and output distributions,

$$C = \max_{P(V)} \sum_V P(V)P(R|V) \, log_2 \frac{P(R|V)}{P(R)} \qquad (1)$$

Figure 2b demonstrates a simple example of a two state system, with input voltages $V_0$ and $V_1$. If there is no overlap between the output distributions $P(R|V_0)P(V_0)$ and $P(R|V_1)P(V_1)$, the capacity is the log (base two) of the number of input states (1 bit). If there is complete overlap, the capacity is 0 bits, as it is impossible to infer V from R. For cases of partial overlap, redundancy needs to be added to the signal to correctly infer the inputs, reducing the effective bits/device. Increasing the number of read states can also increase the capacity for a given number of write states, as seen in Fig. 2c. The capacity achieving input distribution given by equation 1 exhibits an optimal tradeoff between having as many input states as possible and having as little overlap in the output distributions as possible.

Practically, having more read states than write states give additional information in the form of greyscale belief as to which input state they belong. This 'soft information' is currently used by MLC-Flash LDPC encoders to improve inference during belief propagation decoding [7]. Intuitively, analog representations that use the resistance values directly can have higher capacity because they operate at higher granularity (ultimately limited by circuit noise).

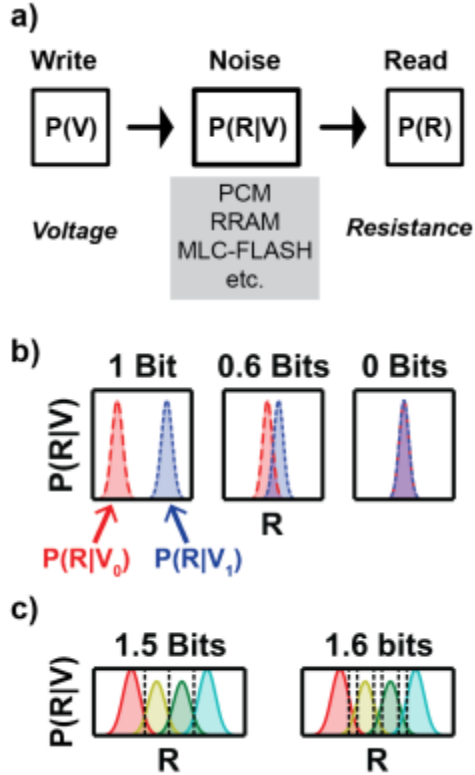

*Figure 2: Storage as a noisy channel. (a) The capacity (equation 1) is determined by the ability to infer a write voltage pulse distribution P(V) from the read resistance distribution P(R). The conditional dependence between the two, P(R|V), is unique for each technology and pulsing scheme. (b) Example of two state system, with input voltages $V_0$ and $V_1$. If there is no overlap between the output distributions (left), the capacity is $log_2$(# write states) = 1 bit. For complete overlap (right), the capacity is 0 bits, as it is impossible to infer anything about the inputs from reading the outputs. For partial overlap (center), redundancy needs to be added to the signal to correctly infer the inputs, reducing the effective bits/device. (c) 'Soft information' increases capacity. For a given number of write states (4), the capacity increases with number of read states (4 left, 7 right). The read resistance is discretized into bins separated by black dotted lines. Even though there are only 4 input states, the extra read states increase the capacity of the system by providing 'soft information', the degree of belief that a read value belongs to each write value. Further read states provide greater granularity of belief values, allowing for easier inference of the input values. Analog codes, such as artificial neural networks, use the actual resistance values, intrinsically benefiting from high granularity.*

**Increased Capacity**

For this study we performed pulsed resistance measurements on 100-device PCM arrays (Fig. 3a). Details of fabrication and characterization of these arrays can be found in previous reports [14,15,16]. The cell is reset to high resistance states via short current pulses that heat the film above its melting temperature and quickly quench to form a resistive amorphous cap. Slow Joule heating above the crystallization temperature anneals the amorphous cap and sets the low resistance state. Multiple resistance levels are achieved by switching between different volumes of resistive amorphous and conducting crystalline phases [3].

To prevent cross-talk and sneak paths in the array, each device is equipped with its own access transistor (1T1R). The gate voltage ($V_{WL}$) on each access transistor limits the current flowing through the cell (Fig. 3a). The gate voltage determines the flow of current to the memory cell and thereby determines the amount of Joule heating (and the local temperature) in the memory cell.

To ensure independent statistics, we apply a pulsing scheme (Fig. 3b) where the cell is initialized to a consistent set state, before applying partial reset pulses to the word line ($V_{WL}$) of varying magnitude and measuring the resulting resistance (R). This facilitates analysis by constraining device behavior to a memoryless zeorth-order markov process.

After collecting 380 trials at each voltage level, we calculate a Gaussian kernel density estimate of the continuous probability density $P(R|V_{WL})$ (Fig. 3c). For simplicity and generality, time-dependent effects of PCM such as resistance drift are not considered here [17]. Similarly, we apply only a single write and read step, even though more robust storage has been demonstrated with read-verify schemes [18]. The results of this investigation can be extended to these modified channel models, and methods for calculating their Shannon capacity have been proposed [19].

Measuring $P(R|V_{WL})$ provides a powerful tool to optimize the write pulse and read bin locations of an associated memory controller. We solve for the capacity-achieving input distribution by maximizing the mutual information over P(V) using the Blahut-Arimoto algorithm [21]. While the algorithm only applies to discrete distributions, we can approximate the capacity of analog channels by sufficiently discretizing P(R|V) such that the capacity is not increased by further discretization (ex. >2000 states).

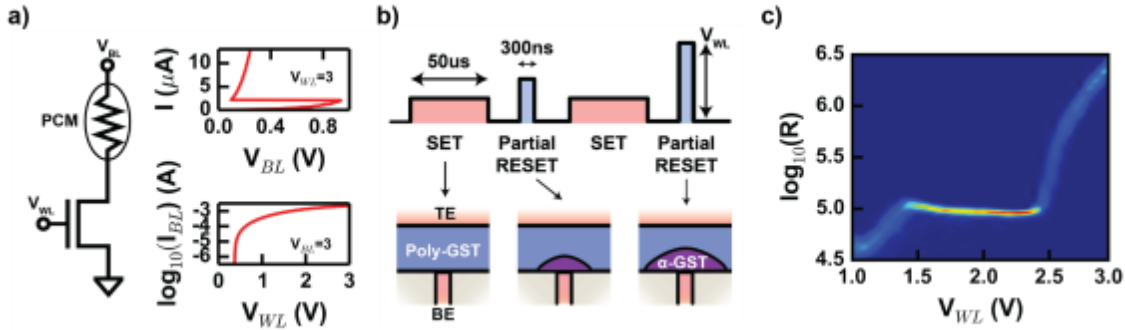

*Figure 3: Measuring P(R|V). (a) Circuit diagram of 10x10 PCM memory array. Devices in the array are individually addressed by applying a wordline voltage ($V_{WL}$) to the gate of the access transistor. The top panel shows characteristic PCM switching behavior for a DC sweep of the bitline voltage ($V_{BL}$) while holding $V_{WL}$ at 3 volts. PCM resistance is modulated by current flow whose magnitude is controlled via applying a large bitline voltage ($V_{BL}$=3V) and pulsing the wordline (bottom panel). Larger wordline voltages create larger currents through the device. (b) Illustration of pulsing scheme. Between each measurement, the cell is consistently set into the low resistance state via a long current pulse that anneals the amorphous capping region (left). The cell is then brought to a higher resistance with a short current pulse that melts the region above the bottom electrode (BE) and quickly cools to form a resistive amorphous cap (middle). Larger current pulses (larger $V_{WL}$), create larger amorphous caps and higher cell resistances (right). (c) Heat map of conditional dependence, $P(R|V_{WL})$, of cell resistance (R) on height of voltage pulse ($V_{WL}$). Density is estimated using Gaussian kernel estimation on 15,200 data points collected between 40 different values of $V_{WL}$. After an initial increase, the amorphous cap reaches a metastable state and only increases in resistance at high values of $V_{WL}$ that have enough energy to melt the entire region and increase the size of the cap.*

The number of nonzero values in the capacity-achieving input distribution is determined by the balance between using as many input states as possible, and reducing overlap of their outputs (Fig. 4a). Interestingly, beyond 12 discrete inputs, no more states are added, as the increased overlap would create a net reduction in mutual information. The output distributions for these 12 inputs partially overlap, demonstrating higher capacity than achievable with totally distinct states. The unequal input state probabilities are unrealistic for most applications, however, making these probabilities uniform only reduces the channel capacity by approximately 5%.

Realistic digital memory controllers have a finite number of read and write states. We model this behavior by marginalizing the continuous probability P(R|V) between discrete read levels, creating a reduced discrete channel. We then use simulated annealing to search for optimal values of the read levels to maximize capacity. Figure 4b demonstrates how the capacity increases with the number of read levels for a given number of write levels. In each case, the capacity monotonically asymptotes as the number of read levels increases, corresponding to the case of an analog read circuit. As the number of allowed input levels increases above 11, the optimal amount for the full analog channel, the discrete capacity asymptotes to the analog capacity of 2.68 bits. This demonstrates that analog memory systems intrinsically operate at peak capacity, assuming circuit noise is small compared to the noise of the memory device itself.

## Energy Efficient Storage

We calculate the energy consumption of each pulse from oscilloscope measurements. The current is transistor limited, indicating that the dynamic resistance of the PCM devices is low and pulses are not strongly affected by parasitic capacitances in the 1T1R structure. Additional power consumed due to line losses are included in simulations of a 1Mb square array with 130 nm wide, 1:1 aspect ratio, Cu wires [20]. We then perform constrained optimization to find the input distributions that maximize capacity per unit energy [21] (Fig. 5a). Since pulses with larger $V_{WL}$ consume more energy, inputs are constrained to lower voltages in the efficient case (Fig. 5b). Although this gives less separable outputs and lower capacities, there is an overall gain in efficiency (nJ/bit). We find an appropriate choice of input pulses can create a 32% reduction in energy/bit for the array.

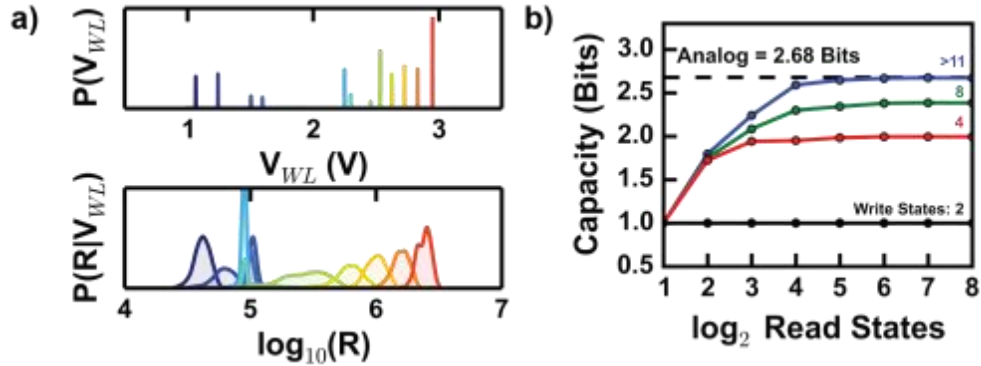

**Figure 4: Analog distributions achieve the highest capacity. (a)** Capacity achieving input distribution, P($V_{WL}$), and corresponding output distribution, P(R). The optimal distributions contain as many input states as possible, while minimizing overlap, resulting in 12 input states. **(b)** Discrete capacity as a function of number of read and write states. Limited by the number of write states, capacity increases with the number of read states due to the creation of 'soft information'. For more than 11 write states, the discrete capacity asymptotes to the analog capacity as the number of read states increases. Thus, error correcting codes that utilize analog circuits and the actual cell resistance values (such as a current summer in an artificial neural network) can achieve the highest possible rates.

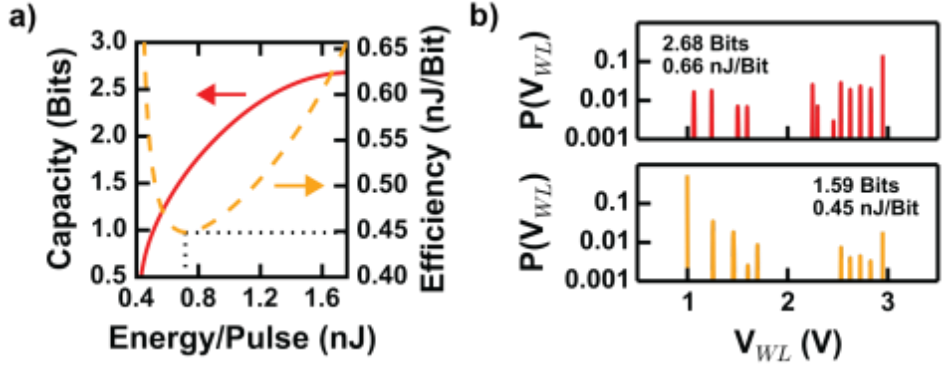

*Figure 5: Energy efficient analog memory system design. (a)* Analog capacity (red) as a function of energy per a pulse measured from individual devices and simulated for a 1Mb array. The constrained optimization shows that capacity decreases with allowed energy, but at a slower rate such as to create a minimum in energy per a bit (yellow). *(b)* Input distributions ($V_{WL}$) for the cases of maximum capacity (red) and maximum efficiency (yellow). The maximum capacity case uses more of the full dynamic range, resulting in more separable states and higher capacity, while the maximum efficiency case uses less high energy consuming states (high $V_{WL}$), resulting in greater efficiency.

## Joint Coding

Finally, we explore an approach to directly store analog-valued information in the resistance values of the PCM devices. While many interesting opportunities exist for exploiting the rich statistics of natural signals, for this preliminary study we restrain ourselves to the canonical case of a Gaussian source signal with an MSE distortion metric. The most naïve approach to store analog values would be employ no coding and directly map source values into voltages (Fig 5a, blue line) and directly provide reconstruction from resistances (Fig 5b, blue line). Such a naïve approach leads to an "effective channel" (conditional probability distribution of reconstruction, $P(\hat{S}|S)$) identical to the original channel ($P(R|V)$). While this is the ideal approach for a Gaussian source with Gaussian channel, MSE distortion, and average power cost [13], the nonlinear PCM device is far from a Gaussian channel. Correspondingly, the naïve implementation exhibits a low SNR (<0dB) at a rate of 1 device/symbol (Fig. 5e, blue dot).

Given that we are working with a Gaussian source and MSE distortion, we would like a parameterized model to learn an encoding and decoding function that minimizes the average cost [22]. We start with a parameterized model,

$$V = F(S) \quad (2)$$
$$\hat{S} = G(R(V)) \quad (3)$$

Where F and G are piecewise nonlinear mappings corresponding to encoder and decoder respectively. We then solve for the mappings that minimize distortion,

$$D^* = \min_{F,G} \iint (\hat{S} - S)^2 P(\hat{S}|S) P(S) \, d\hat{S} dS \quad (4)$$

As we restrict ourselves to a single input and output dimension, discretized at 1000 values, we are able to solve for the parameters of F and G via complete enumeration.

To better understand the effects of coding, we examine the results of having both an encoder and decoder (Fig. 5, red line) as well as the encoder (Fig. 5, yellow) and decode (Fig. 5, green) on their own. Interestingly, the mappings are non-monotonic, reflecting the channel itself. The encodings (Fig. 5a) in particular show a sharp transition, as voltages ~1.5V are better choices for large source values than ~2.3V due to their sharper P(R|V) distributions. The resulting effective channel of these optimal encodings/decodings (Fig. 5d), is not surprisingly linear, mimicking the

zero mean correspondences of a Gaussian channel. We observe that coding reduces the distortion, with joint encoding-decoding increasing SNR more than either in isolation. Even with this simple encoding/decoding scheme (amounting to a single dimensional lookup table) joint coding with these devices achieves a similar SNR to a separate coding schemes of arbitrary complexity at 2 bits per a device (Fig. 5d, dotted line). From another perspective, the performance is equal to a system with capacity achieving channel coding (2.68 bits) and simple vector quantization as source coding (Fig. 5d, dashed line). While we have only examined the case of single dimensional encodings here, the design for analog encoders/decoders is rich for higher dimensions and other compression ratios, making it a promising field for the development of coding circuitry that is low in complexity, energy, and latency

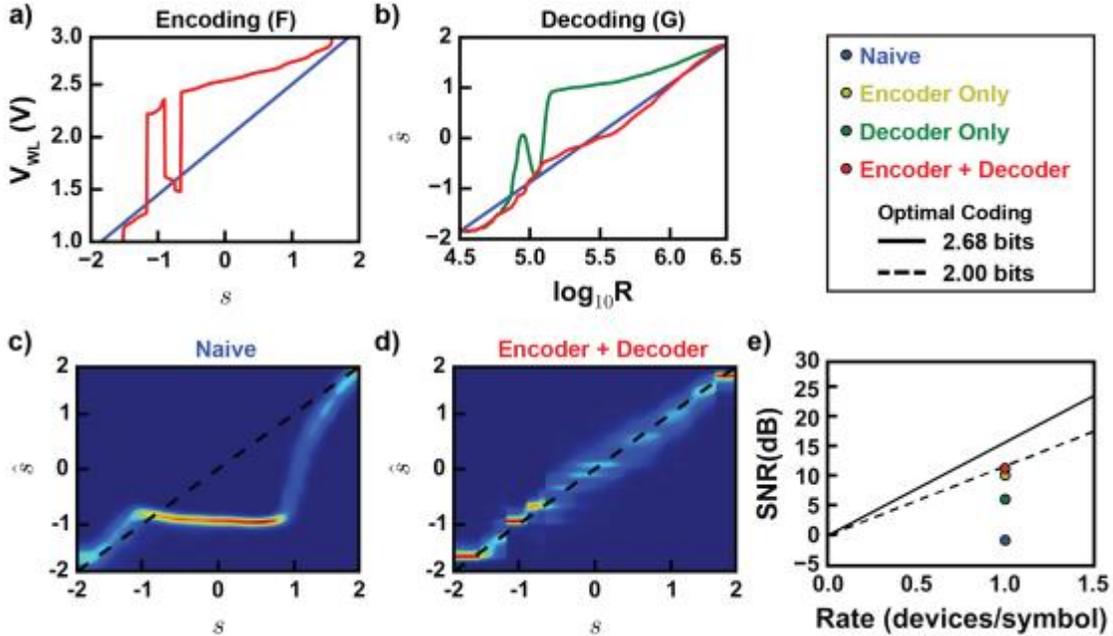

*Figure 6: Optimal joint coding of a Gaussian source.* *1:1 joint encoding **(a)** and decoding **(b)** strategies for storing a Gaussian source with mean 0 and variance 1 directly in the resistance of PCM devices on a symbol-by-symbol basis (block length=1). In the naïve case (blue), the source is encoded linearly into voltage, $V_{WL}$, and decoded linearly from resistance, R. This results in a conditional dependence **(c)** of the reconstruction on the source, $P(Ŝ|S)$, that resembles the device characteristics in Fig. 3c. Nonlinear mappings for the encoder only (yellow), decoder only (green), and both (red, optimal) are solved for by minimizing the MSE distortion (eqs, 2, 3, 4). The nonmonotonicity of the optimal encoder arises because $P(R|V_{WL})$ in Fig. 3c is slightly nonmonotonic itself, resulting in $P(Ŝ|S)$ that rises linearly **(d)**. **(e)** Rate (devices/symbol) versus distortion (SNR in decibels) comparison of symbol-by-symbol joint coding strategies (colored circles) and optimal separate coding strategies in the limit of infinite block length (black lines). From Fig. 4b, we see PCM with an analog ECC can reach a capacity of 2.68 bits (solid line), and more realistic discrete systems with 8 read and write states are close to 2 bits (dotted line). For additional comparison, we show R(D) for C=2.68 bits with optimal channel coding but source coding with a block length of 1 (scalar quantization, dashed line). Despite the reduced complexity (two lookup tables) and block length (1) of joint coding, it achieves similar performance to 2 bit systems with optimal source and channel coding and 2.68 bit systems with optimal channel coding. Further improvements are likely for systems that can learn higher dimensional nonlinear mappings, such as artificial neural networks, with longer block lengths (k:k) or different rates (k:m).*

## Discussion

In this paper, we have explored a single approach to storing analog-valued media in emerging analog-valued devices, and found that analog coding strategies have the potential to create robust storage with relatively low complexity. Designing systems to match the statistics of the source, channel, and task-at-hand provides unique opportunities not captured by systems that try to enforce determinism. However, there is no free lunch and what such systems possibly gain in efficiency they lose in generality.

Universal memory design has been a great boon to the development of digital technology, and designing memory systems for specific tasks has many unexplored risks in terms of backwards compatibility. That said, with the rise of machine learning applications, the nature of computation is changing and many of the data centric applications of the present and future do not require perfect data retrieval. With the data deluge faced by internet companies, it is often not the storage of data that is important, but the storage of information which performs well when retrieved for statistical machine inference or human inference (i.e. perception of audio and video). While losing some generality, representations that reduce the redundancy of signals to match the statistics of storage media and applications may be more efficient than universal representations.

It is interesting to note that from the perspective of neuroscience, neural systems are faced with a very similar task. Given the statistical redundancy of natural stimuli, and the stochasticity of perception and signal transmission, efficient coding strategies must be devised to transmit compressed representations of stimuli that are intrinsically robust to neural noise and resource efficient [23, 24].

Finally, the results presented here represent only a single model applied to a single type of signal and device. The future design space to explore in this field is extremely rich. From the experimental perspective, other devices (RRAM, CBRAM, and MLC-Flash) can exhibit dramatically different behavior. Different pulsing schemes can produce different types of statistics. For example, iterative pulsing introduces rich conditional dependencies through time that allow are currently exploited by multibit memory systems and present an interesting opportunity from the perspective of modeling. Cell-to-cell dependencies within the array are also a fact of life that must usually be avoided through hand-engineered systems. Measuring these statistical dependencies can enable codes to be adaptive to the flaws and statistics of individual chips, both at the factory during test and over the lifetime of use, improving the reliability and uniformity of performance.

From an algorithmic perspective, many powerful analog-valued high-dimensional models such as artificial neural networks exist, leaving a rich space to explore for learning high-dimensional parameterized encodings and compression. Since such models have been able to successfully utilize the statistics of natural media such as images [25] and sound [26], it is promising that they will also perform well for encoding such media in analog-valued devices, and recent research has yielded impressive results [27]. Efficient integrated circuit implementations of such circuits are also an active area of investigation [28] lending further weight to the idea of analog-valued ECCs that can learn on-chip dependencies and adapt to nonstationary statistics as devices age. Another active area of research is in using analog-valued variable resistors as synaptic elements in such neural circuits [29], leading to the possibility that storage and error-correction could someday be integrated within a single memory substrate.

## Acknowledgments
This work was supported in part by SONIC, one of six centers of STARnet, a Semiconductor Research Corporation program sponsored by MARCO and DARPA, and by member companies of the Stanford Non-Volatile Memory Technology Research Initiative (NMTRI).

**Author Contributions**

J.H.E. performed all experiments and simulations and wrote the manuscript. J.H.E., H.S.P.W., and B.A.O. conceived of the project. S.B.E. assisted with experimental setup and design. S.K., M.B., C.L., and H.L.L., provided the test wafer of memory devices and helpful discussions.

**Competing Financial Interests**

The authors claim no competing financial interests in conducting this research.

**Methods**

------------

Pulsed experiments were conducted on a Cascade probe station. Wafers, provided by IBM, were contacted by a custom probe card and connected to a Keithley 700B Switch Matrix, enabling access to 100 unique devices. Voltage pulses were produced by a Agilent 81110A Pulse Generator. 'Set' pulses applied 3V to the wordline and bitline with a 500ns rise time, 5000ns width, and 500ns tail. Partial reset pulses applied 3V to the bitline and a variable voltage to the wordline with 5ns rise time, 50ns width, and 5ns fall time. Cell resistances were measured with a Agilent 4155C Semiconductor Parameter Analyzer, applying 3V to the wordline and 0.1V to the bitline. All equipment was controlled via home-built python software utilizing PyVisa and wxPython (available at http://www.github.com/jesseengel/PythonProbestation).

Capacity calculations and rate-distortion simulations were conducted in python using Numpy. The implementation of the Blahut-Arimoto algorithm was adapted from matlab code by Kenneth Shum (http://home.ie.cuhk.edu.hk/~wkshum/wordpress/?p=825). Estimates of P(R|V) were calculated from experimental data using gaussian kernel density estimation.

**References**

[1] Evans, D. The Internet of Things: How the Next Evolution of the Internet Is Changing Everything. (2011). At
<http://www.cisco.com/web/about/ac79/docs/innov/IoT_IBSG_0411FINAL.pdf>

[2] MacKay, D. J. C. Information Theory, Inference, and Learning Algorithms. (Cambridge University Press, 2003).

[3] Wong, H.-S. P. et al. Phase Change Memory. Proc. IEEE 98, 2201–2227 (2010).

[4] Wong, H.-S. P. et al. Metal–Oxide RRAM. Proc. IEEE 100, 1951–1970 (2012).

[5] Shannon, C. E. A Mathematical Theory of Communication. Bell Syst. Tech. J. 27, 379–423 (1948).

[6] Wang, Z., Bovik, A. C., Sheikh, H. R. & Simoncelli, E. P. Image quality assessment: from error visibility to structural similarity. *IEEE Trans. Image Process.* **13**, 600–12 (2004).

[7] Motwani, R., Kwok, Z. & Nelson, S. Low Density Parity Check (LDPC) Codes and the Need for Stronger ECC. in *Flash Memory Summit* (2011). at
<http://www.flashmemorysummit.com/English/Collaterals/Proceedings/2011/20110810_T2A_Nelson.pdf>

[8] MacKay, D. J. C. & Neal, R. M. Near Shannon limit performance of low density parity check codes. Electron. Lett. 33, 457 (1997).

[9] Jonghong Kim & Wonyong Sung. Rate-0.96 LDPC Decoding VLSI for Soft-Decision Error Correction of NAND Flash Memory. *IEEE Trans. Very Large Scale Integr. Syst.* **22**, 1004–1015 (2014).

[10] Mansour, M. M. & Shanbhag, N. R. A 640-Mb/s 2048-Bit Programmable LDPC Decoder Chip. *IEEE J. Solid-State Circuits* **41**, 684–698 (2006).

[11] Micheloni, R. et al. A 4Gb 2b/cell NAND Flash Memory with Embedded 5b BCH ECC for 36MB/s System Read Throughput. *Solid-State Circuits Conf. 2006. ISSCC 2006. Dig. Tech. Pap. IEEE Int.* **32**, 497–506 (2006).

[12] Kostina, V. & Verdu, S. Lossy joint source-channel coding in the finite blocklength regime. In 2012 IEEE International Symposium on Information Theory Proceedings 1553–1557 (IEEE, 2012).

[13] Gastpar, M., Rimoldi, B. & Vetterli, M. To code, or not to code: lossy source-channel communication revisited. *IEEE Trans. Inf. Theory* **49**, 1147–1158 (2003).

[14] Breitwisch, M. et al. Novel Lithography-Independent Pore Phase Change Memory. in *2007 IEEE Symposium on VLSI Technology* 100–101 (IEEE, 2007).

[15] Close, G. F. et al. Device, circuit and system-level analysis of noise in multi-bit phase-change memory. in *2010 International Electron Devices Meeting* 29.5.1–29.5.4 (IEEE, 2010).

[16] Eryilmaz, S. B. et al. Experimental Demonstration of Array-level Learning with Phase Change Synaptic Devices. in *2013 IEEE International Electron Devices Meeting* **25**, 1–4 (IEEE, 2013).

[17] Papandreou, N. et al. Drift-Tolerant Multilevel Phase-Change Memory. in *2011 3rd IEEE International Memory Workshop (IMW)* 1–4 (IEEE, 2011).

[18] Nirschl, T. et al. Write Strategies for 2 and 4-bit Multi-Level Phase-Change Memory. in *2007 IEEE International Electron Devices Meeting* 461–464 (IEEE, 2007).

[19] Lastras-Montano, L. A., Franceschini, M., Mittelholzer, T. & Sharma, M. Rewritable storage channels. in *2008 International Symposium on Information Theory and Its Applications* 1–6 (IEEE, 2008).

[20] Liang, J., Yeh, S., Wong, S. S. & Wong, H.-S. P. Effect of Wordline/Bitline Scaling on the Performance, Energy Consumption, and Reliability of Cross-Point Memory Array. *ACM J. Emerg. Technol. Comput. Syst.* **9**, 1–14 (2013).

[21] Arimoto, S. An algorithm for computing the capacity of arbitrary discrete memoryless channels. *IEEE Trans. Inf. Theory* **18**, 14–20 (1972).

[22] Akyol, E., Rose, K. & Ramstad, T. Optimal mappings for joint source channel coding. in *IEEE Information Theory Workshop* 2010 (ITW 2010) 1–5 (IEEE, 2010).

[23] Karklin, Y. & Simoncelli, E. Efficient coding of natural images with a population of noisy Linear-Nonlinear neurons. *NIPS* 1–9 (2011).

[24] Varshney, L. R., Sjöström, P. J. & Chklovskii, D. B. Optimal information storage in noisy synapses under resource constraints. *Neuron* **52**, 409–23 (2006).

[25] Alex Krizhevsky, Ilya Sutskever, G. E. H. Imagenet classification with deep convolutional neural networks. in *NIPS* 1106–1114 (2012).

[26] Geoffrey Hinton, Li Deng, Dong Yu, George Dahl, Abdel-rahman Mohamed, Navdeep Jaitly, Vincent Vanhoucke, Patrick Nguyen, Tara Sainath, B. K. Deep Neural Networks for Acoustic Modeling in Speech Recognition. *IEEE Signal Process. Mag.* **29**, 82–97 (2012).

[27] Toderici, G., Vincent, D., Johnston, N., Hwang, S. J., Minnen, D., Shor, J., & Covell, M., Full Resolution Image Compression with Recurrent Neural Networks., *CoRR*, abs/1608.05148, (2016).

[28] Neftci, E., Das, S., Pedroni, B., Kreutz-Delgado, K. & Cauwenberghs, G. Event-driven contrastive divergence for spiking neuromorphic systems. *Front. Neurosci.* **7**, 272 (2013).

[29] Kuzum, D., Yu, S. & Wong, H.-S. P. Synaptic electronics: materials, devices and applications. *Nanotechnology* **24**, 382001 (2013).